\documentclass[fleqn,usenatbib]{mnras}

\usepackage{newtxtext,newtxmath}

\usepackage[T1]{fontenc}

\DeclareRobustCommand{\VAN}[3]{#2}
\let\VANthebibliography\thebibliography
\def\thebibliography{\DeclareRobustCommand{\VAN}[3]{##3}\VANthebibliography}

\usepackage{graphicx}	
\usepackage{subcaption}
\usepackage{amsmath}	
\usepackage{bm}

\title[Information content of WL field]{
Map-based cosmology inference with weak lensing -- information content and its dependence on the parameter space}
\author[S. S. Boruah et al.]{Supranta S. Boruah$^{1}$\thanks{Contact e-mail: \href{mailto:ssarmabo@email.arizona.edu}{ssarmabo@email.arizona.edu}}, Eduardo Rozo$^{2}$\thanks{Contact e-mail: \href{mailto:erozo@email.arizona.edu}{erozo@email.arizona.edu}}
\\
$^{1}$ Department of Astronomy and Steward Observatory, University of Arizona, 933 N Cherry Ave, Tucson, AZ 85719, USA \\
$^{2}$Department of Physics, University of Arizona, 1118 E. Fourth Street, Tucson, AZ, 85721, USA
}

\date{Accepted XXX. Received YYY; in original form ZZZ}

\pubyear{2022}
\begin{document}
\label{firstpage}
\pagerange{\pageref{firstpage}--\pageref{lastpage}}
\maketitle

\begin{abstract}
Field-level inference is emerging as a promising technique for optimally extracting information from cosmological datasets.  Indeed, previous analyses have shown field-based inference produces tighter parameter constraints than power spectrum analyses. However, estimates of the detailed quantitative gain in constraining power differ.  Here, we demonstrate the gain in constraining power depends on the parameter space being constrained.  As a specific example, we find that field-based analysis of an LSST Y1-like mock data set only marginally improves constraints relative to a 2-point function analysis in $\Lambda$CDM, yet it more than doubles the constraining power of the data in the context of $w$CDM models. This effect reconciles some, but not all, of the discrepant results found in the literature. Our results demonstrate the importance of using a full systematics model when quantifying the information gain for realistic field-level analyses of future data sets.
\end{abstract}


\section{Introduction}\label{sec:intro}

Current lensing analyses typically rely on 2-point functions \citep{Hikage2019, Heymans2021, DESY3_2022}. 
However, 2-point analyses are sub-optimal due to the highly non-Gaussian nature of the late-time density field. Indeed, one can extract additional cosmological information by supplementing 2-point function measurements with non-Gaussian summary statistics \citep{Takada2003, Kilbinger2005}, e.g. peak counts \citep{Liu2015, HarnoisDeraps2021, Zurcher2022}, one-point PDFs \citep{Thiele2020, Boyle2021}, wavelet transforms \citep{Cheng2020, Cheng2021, Ajani2021}, and Minkowski functionals \citep{Kratochvil2012, Petri2013}.

Field-level inference \citep{Jasche2013, Wang2014, Modi2018} is a new approach in which one forward-models the cosmology-dependent density field of the Universe as constrained by the data. A field-based inference approach is fully optimal at any given scale: it automatically and self-consistently incorporates {\it all} summary statistics up to the recovered scale. For this reason, it has been proposed to model a broad range of observables, including weak lensing \citep{Porqueres2021, Porqueres2022, Fiedorowicz2022, Fiedorowicz2022a, Boruah2022}, CMB lensing \citep{Millea2019, Millea2020, Millea2021}, peculiar velocities \citep{Boruah2022_velocity, PrideauxGhee2022, Bayer2022}, and galaxy clustering \citep{Ramanah2019,Dai2022}. Although numerically challenging, steady progress in numerical techniques \citep{Modi2021, Li2022, Modi2022, Dai2022} is helping realize the potential of this new technique.

While there is consensus in the literature that field-based inference leads to tighter parameter constraints than  2-point analyses, there are also significant differences in the detailed quantitative measure of this improvement. \citet{Leclercq2021} found that field-based inference leads to massive improvement in parameter constraints over 2-pt function analysis, even for only mildly non-Gaussian fields. Similarly, \citet{Porqueres2022, Porqueres2023} found large gains for a field-level cosmic shear analysis.  By contrast, \citet{Boruah2022} found field-based inference results in only modest improvements for cosmic shear analyses.  In light of these differences, we have set out to examine the information gain from field-level inference of weak lensing data in more detail.

\begin{figure*}
\centering
\begin{subfigure}{.5\textwidth}
  \centering
  \includegraphics[width=\linewidth]{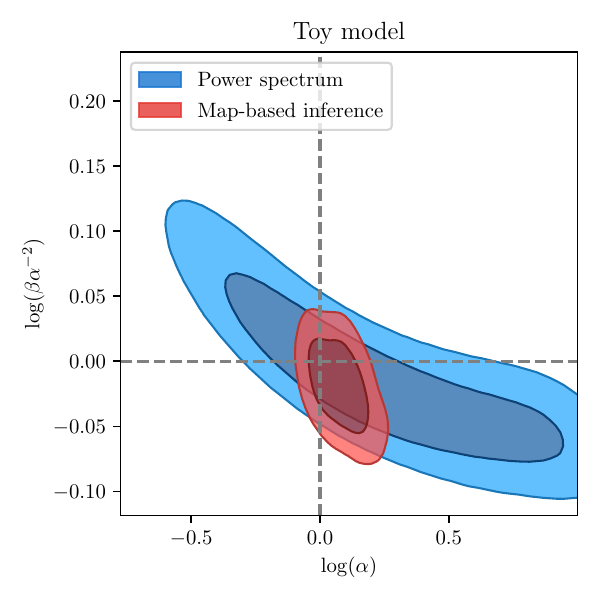}
\end{subfigure}%
\begin{subfigure}{.5\textwidth}
  \centering
  \includegraphics[width=\linewidth]{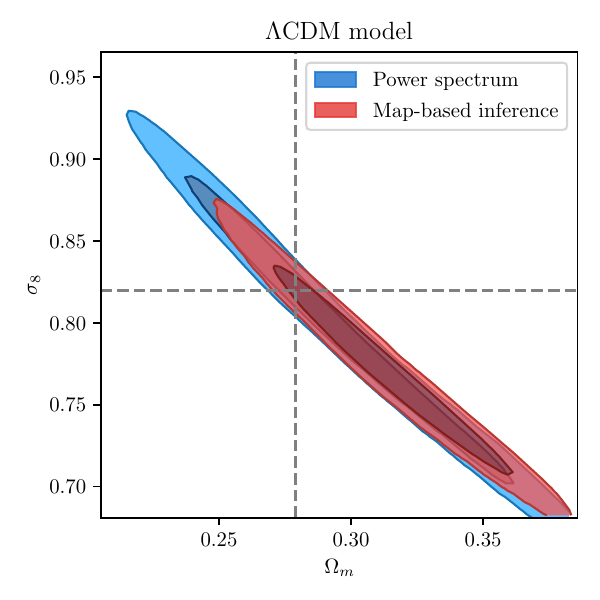}
\end{subfigure}
\caption{Comparison of the constraints obtained using field-based inference and power spectrum analysis for the toy model  described in section \ref{sec:linear_toy_model} ({\it left}), and a $\Lambda$CDM cosmological model ({\it right}). We use the same observed data vector --- i.e. the noisy realization of the observed shear field for the two panels. We plot $\beta / \alpha^2$ on the $y$ axis for the toy model to account for the strong degeneracy between these two parameters. We see that field-based inference dramatically improves parameter constraints in the $\alpha$-$\beta$ toy model, but have only a modest impact on cosmological posteriors \citep{Boruah2022}. That is, the gains due to field-based inference methods relative to 2-point analyses depend on the parameter space under consideration.
}
\label{fig:toy_model_contours}
\end{figure*}

\section{Formalism}\label{sec:formalism}

We model the convergence field as a lognormal random field.  Lognormal fields are commonly used to approximate non-Gaussian density and convergence fields in cosmological applications \citep{Coles1991, Jasche2010, Clerkin2017, Xavier2016}. Throughout this paper, we perform our analysis at a pixel scale of 10 arcminutes.  This is sufficiently large for the lognormal distribution to provide a reasonable description of the underlying convergence field \citep{ Xavier2016, Clerkin2017, Friedrich2020}. We do not consider smaller scales to avoid having to model baryonic feedback, which is expected to significantly impact the matter density distribution at higher resolution \citep[e.g, ][]{Eifler2015, Huang2019, Osato2021}.

When modelled as a lognormal variable, $\kappa$ is related to a Gaussian variable $y$ via
\begin{equation}
    \kappa = e^{y} - \lambda,
\end{equation}
where $\lambda$ is called the shift parameter. The shift parameter denotes the minimum value that $\kappa$ can take, and directly impacts the non-Gaussian features of the resulting convergence field. The mean of the $y$-field is chosen so as to enforce the condition that the $\kappa$ field has a zero mean. We use the perturbation theory code {\sc cosmomentum} \citep{Friedrich2018, Friedrich2020} to calculate the cosmology-dependent shift parameters. For further details on lognormal fields, we refer the reader to \citet{Boruah2022}.

We use the field-level analysis pipeline of \citet{Boruah2022} to analyze synthetic weak lensing data generated from a lognormal convergence map.  To create the synthetic data, we assume the redshift distribution forecasted for LSST-Y1 in \citet[DESC-SRD,][]{DESC_SRD}. We then analyze the synthetic data using two different models: {\it (i)} a two-parameter toy model presented in Section \ref{sec:linear_toy_model}, and {\it (ii)} a cosmological model in which the power spectrum and the shift parameters are determined by the underlying cosmological parameters. Following \citet{Leclercq2021}, the toy-model analysis of section~\ref{sec:linear_toy_model} is non-tomographic.  The cosmological analysis of section~\ref{sec:lcdm_vs_wcdm} assumes the data is binned into 4 tomographic bins.
\section{Toy model with scaling parameters}\label{sec:linear_toy_model}

\citet{Leclercq2021} used a two-parameter log-normal toy model to demonstrate that field-based analyses can dramatically outperform standard 2-point approaches. This result is apparently in tension with that of \citet{Boruah2023}, who find only marginal improvements in a $\Lambda$CDM cosmology.  To resolve this apparent discrepancy, we analyzed a synthetic data set using two different models: a toy model similar to the one used by \citet{Leclercq2021}, and the standard $\Lambda$CDM model. Our toy model is constructed so that our fiducial toy model exactly matches our fiducial $\Lambda$CDM model.

Our fiducial model is a flat $\Lambda$CDM universe with $\Omega_{\rm m}=0.279$, $\sigma_8=0.82$, $\Omega_{\rm b} = 0.046$, $h=0.7$, $n_{\rm s}=0.97$.  This choice defines the power-spectrum $C_y(\ell)$ and the shift parameter $\lambda$ of the lognormal random field $\kappa$, where $y=\ln(\kappa - \lambda)$, and $y$ is a Gaussian random field.  Our toy model depends on two parameters $\alpha$ and $\beta$ that rescale: 1) the power-spectrum $C_y$; or 2) the shift parameter $\lambda$.   These rescalings are defined via
\begin{align}
    \log C_y(\ell) &\rightarrow \alpha \times \log C_y(\ell) \\
    \lambda &\rightarrow \beta \times \lambda.
\end{align}
For simplicity, we refer to this toy model as the $\alpha$-$\beta$ model, with $\alpha=\beta=1$ corresponding to our fiducial model.  As in \citet{Leclercq2021}, we restrict our analysis to a single tomographic redshift bin, for which we adopt the expected redshift distribution of source galaxies for the LSST-Y1 data set.

We produce a lognormal realization of the fiducial model, which we then analyze using the field-based inference framework of \citet{Boruah2022}.  We perform our analysis both in the toy $\alpha$-$\beta$ model and the $\sigma_8$--$\Omega_{\rm m}$ parameter space. Both analyses rely on the same noisy shear map as the data vector.

Figure~\ref{fig:toy_model_contours} compares the posteriors for the $\alpha$--$\beta$ model (left) and the $\Lambda$CDM model (right).  Red and blue contours correspond to posteriors from a field-based (red) and a power spectrum based (blue) analysis.  Evidently, field-based inference dramatically improves parameter constraints in our toy model, but has only a modest impact on the posteriors in the $\sigma_8$--$\Omega_{\rm m}$ space.  This demonstrates that: 1) despite being superficially different, the results of \citet{Leclercq2021} and \citet{Boruah2022} are fully consistent with each other; and 2) the amount of information gained from field-based inference depends on the parameter space of interest.

We can readily understand the difference in gains between the two parameters spaces as follows.  In the $\alpha$--$\beta$ toy model, the 1-point and 2-point functions of the field vary in nonphysical and largely independent ways. However, in the real Universe, the power spectrum and the 1 pt PDF are determined by the same physics and therefore contain correlated information. 

\begin{figure}
    \centering
    \includegraphics[width=\linewidth]{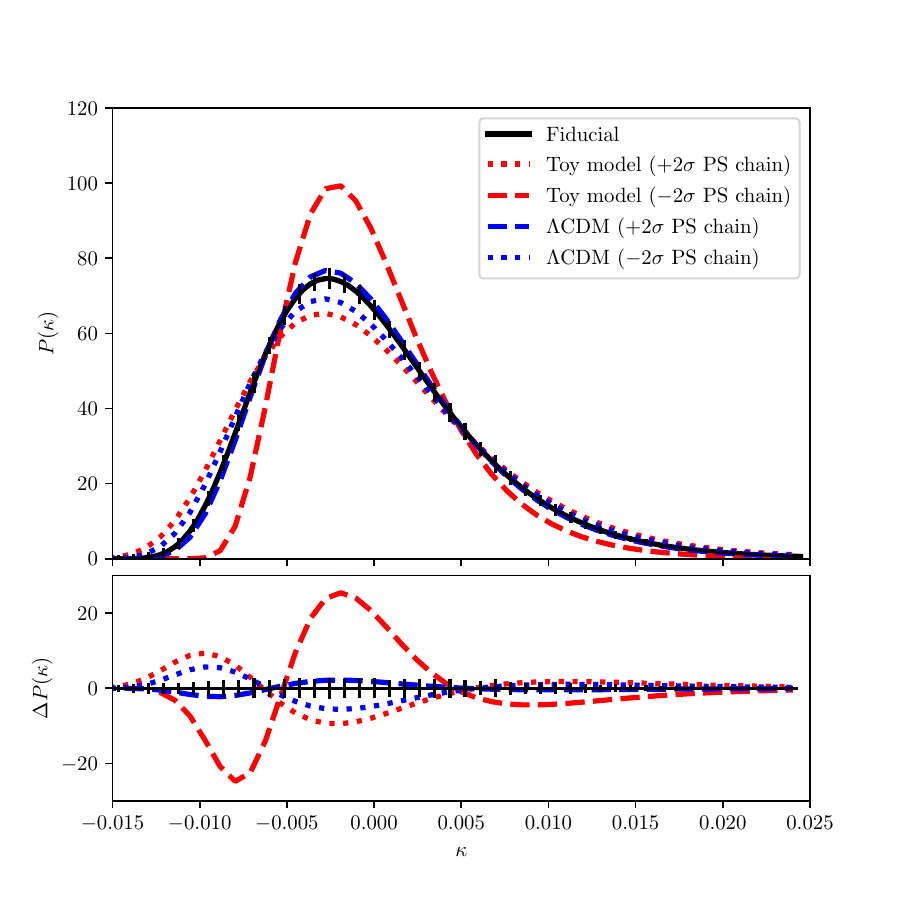}
    \caption{Comparison of the 1-point distributions of models that are $2\sigma$ away from the fiducial value in the 2-point posterior analyses for both the $\alpha$--$\beta$ (red dashed and dotted lines) and $\sigma_8$--$\Omega_{\rm m}$ (blue dashed and dotted lines) parameter spaces. The bottom panel shows the differences between the 1-point distributions from that of the fiducial model.  The error bars in the bottom panel are the noise in the measured 1-point distribution.  Note that the differences in the 1-point distribution are highly significant in the case of the $\alpha$--$\beta$ parameter space, but only marginally significant in the $\sigma_8$--$\Omega_{\rm m}$ space. 
    }
    \label{fig:pdf_sensitivity}
\end{figure}

To demonstrate this, we select models from the 
power spectrum posteriors in each of the two parameter space we considered. The models selected are exactly $2\sigma$ away from the fiducial model and along the degeneracy direction of the 2-point posterior in each space. Figure \ref{fig:pdf_sensitivity} compares the 1-point function for each of these models.  We see that the difference between the 1-point function for each of these models and that of the fiducial model is many times larger in the $\alpha$--$\beta$ parameter space than in the $\sigma_8$--$\Omega_{\rm m}$ space.  Moreover, the differences in the 1-point functions in the $\sigma_8$--$\Omega_{\rm m}$ parameter space are comparable to the cosmic variance errors, explaining why field-based inference results in only marginal gains relative to the 2-point posterior.  In short, the reason the toy-model of \citet{Leclercq2021} results in large gains is because it allows for an unphysical de-correlation of the information content of the 1- and 2-point functions of the convergence field.

\section{Implications for Cosmological Inference}\label{sec:lcdm_vs_wcdm}

\begin{figure*}
    \centering
    \includegraphics[width=\linewidth]{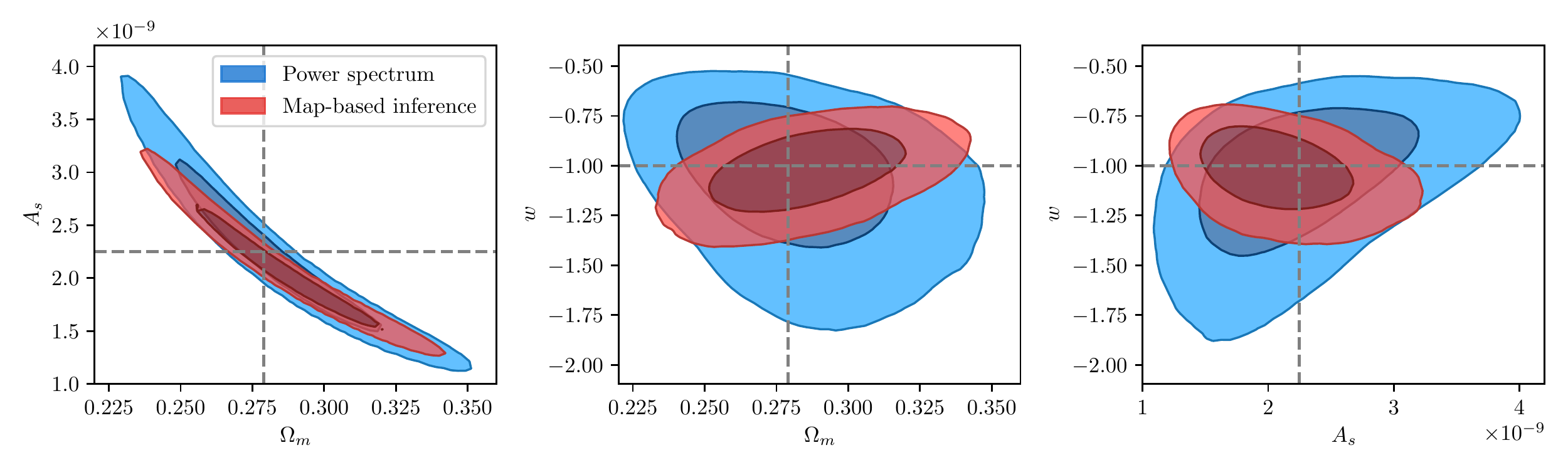}
    \caption{Comparison of the cosmological constraints with power spectrum analysis (blue) and map-based inference (red) for $w$CDM parameter space. We find that map-based inference leads to much stronger constraints than a power spectrum based analysis.  This is in contrast to our findings within the context of $\Lambda$CDM,
    where field-based inference resulted in only modest improvements. }    
    \label{fig:wcdm_results}
\end{figure*}

We have seen that that the choice of parameter space impacts the relative gain of field-based inference methods relative to traditional 2-point analyses. This raises the question: are there other cosmological parameters for which the gain in cosmology constraints is large? Here, we compare cosmological constraints in $w$CDM models from cosmic shear as derived from field-based and power spectrum analyses. In contrast to the previous section, we perform a tomographic analysis with 4 redshift bins, each containing the same number of galaxies. The redshift distribution of the bins is based on the expected LSST-Y1 redshift distributions.  The source density is set to 10 galaxies/arcmin$^2$.

Figure~\ref{fig:wcdm_results} summarizes our results.  The figure demonstrates that a field-based approach significantly improves parameter constraints relative to the standard 2-point analysis in a $w$CDM cosmology. We quantify the improvement using the covariance matrix of the posterior.  Specifically, we define the figure-of-merit 
\begin{equation}
    \text{FoM}_{ij} = \frac{1}{\sqrt{\text{det}(\text{Cov}[\theta_i, \theta_j])}},
\end{equation}
where, $\text{Cov}[\theta_i, \theta_j]$ denotes the covariance matrix of the parameters $\theta_i$ and $\theta_j$ as computed from the MCMC posterior samples. We find that field-based inference leads to an improvement in the figure of merit by a factor of 2.2, 2.2, and 2.5 times in the $\Omega_{\rm m}$--$A_{\rm s}$, $\Omega_{\rm m}$--$w$ and $A_{\rm s}$--$w$ subspaces respectively. These improvements are particularly noteworthy in that the cosmological information content of the shear power spectrum begins to saturate at $\approx 10$ arcmin scales \citep{Kayo2013, Boruah2023}.  That is, field-based analyses are a powerful complement to efforts centered on improving small scale modeling.

\begin{figure*}
    \centering
    \includegraphics[width=\linewidth]{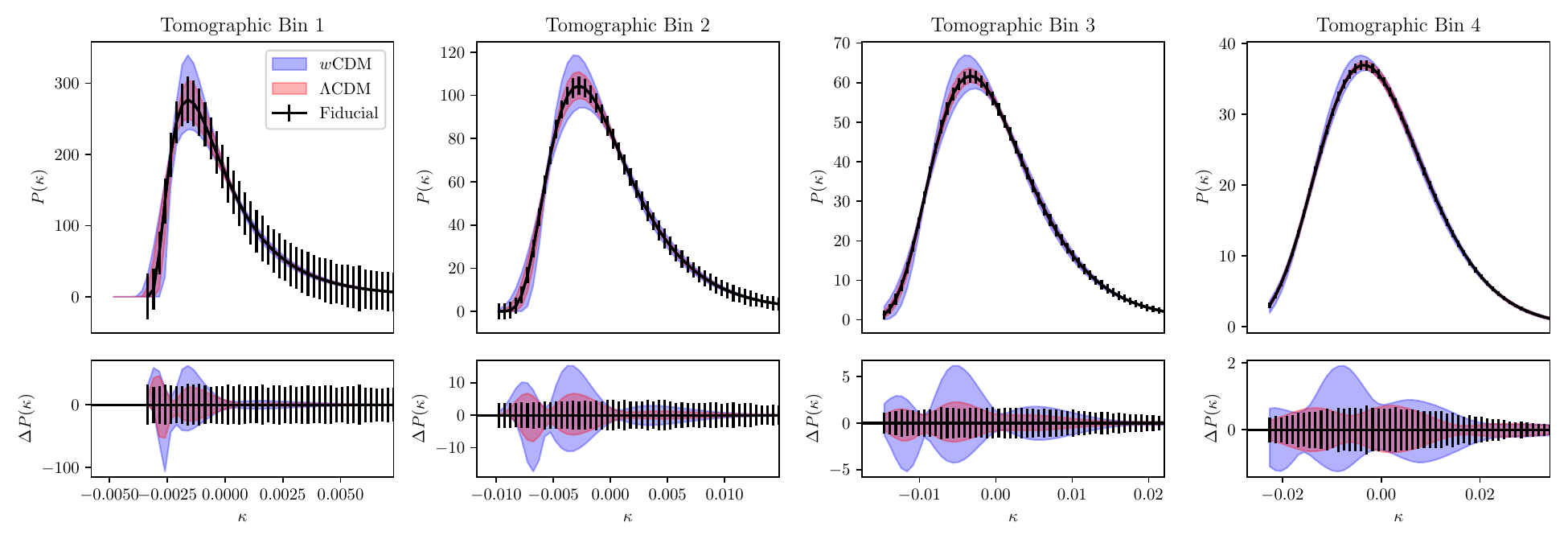}
    \caption{Spread in the 1-point function calculated for the cosmological parameters drawn from the power spectrum posterior for a $\Lambda$CDM (red) and a $w$CDM analysis (blue). The black bars show the expected statistical error in the recovered distributions including shape noise. The differences in the posterior predictions for the 1-point distributions are larger than the observational errors in $w$CDM, but smaller in $\Lambda$CDM. Consequently, field-based inference leads to large improvements in parameter constraints in the context of $w$CDM, but only modest improvements in $\Lambda$CDM.}
    \label{fig:lcdm_wcdm_1pt_func}
\end{figure*}

As in section~\ref{sec:linear_toy_model}, the additional information in the field-based inference analysis comes from the 1-point function.  This is illustrated in Figure \ref{fig:lcdm_wcdm_1pt_func}.  There, we compare: 1) the spread in the predicted 1-point functions obtained by sampling the power-spectrum analysis posterior; and 2) the observational uncertainties in the 1-point distribution.  This comparison is done both for $\Lambda$CDM and $w$CDM posteriors, and each of the four tomographic bins. We see that the spread in the one-point function within the $\Lambda$CDM chain is less than or comparable to the statistical noise in the one-point function measurement. On the other hand, the spread in the predicted 1-point distributions from the $w$CDM power spectrum posterior is broader than observational uncertainties. Consequently, the 1-point distribution function adds significant information to the 2-point analysis for $w$CDM models.  Conversely, a measurement of the 1-point distribution adds little information within the context of a $\Lambda$CDM analysis.


Our results are in tension with those of \citet{Porqueres2022} and \citet{Porqueres2023}, who report large gains from field-based inference in a $\Lambda$CDM cosmology. Barring numerical issues/bugs in one or both of these codes, this discrepancy can only be resolved by the differences in the forward models. The convergence field in \citet{Porqueres2023} is calculated using 2LPT simulations plus ray tracing, whereas we rely on an approximate log-normal model. However the lognormal model provides a good description of the convergence field at the current resolution \citep[e.g,][]{Xavier2016, Clerkin2017, Fiedorowicz2022} and therefore the massive difference between the two results is surprising. Understanding the sources of this discrepancy is outside the scope of this work, but the difference highlights the need for more extensive testing and detailed comparisons between different field-level inference codes. 
In this context we note that in \citet{Boruah2022} we have verified that our posteriors match the analytic expectation when using a Gaussian random field model.     

\section{Summary and Discussion}\label{sec:conclusion}

 We used the lognormal model to study the relative information content from field-based and 2-point analyses of the convergence field.  We confirm the finding that field-based parameter posteriors are significantly tighter than those of the corresponding 2-point analysis in the case of the \citet{Leclercq2021} toy model.  However, we have also demonstrated that the relative gains of field-based inference depend on the specific parameter space being investigated.  In particular, we have found field-based inference leads to modest gains in $\Lambda$CDM, but large gains in $w$CDM. These improvements are driven by the information content in the 1-point distribution of the convergence field.  

It is important to note that in this analysis we have not considered systematic effects. As we saw in section \ref{sec:linear_toy_model}, the constraining power depends on the parameter space considered. Therefore, the addition of systematic parameters to the model will impact our conclusions regarding the impact of field-based inference on cosmological posteriors.  That said, several studies in the literature have shown that non-Gaussian information can improve constraints on systematics parameters such as photo-$z$ biases \citep{Jasche2012, Tsaprazi2023} and intrinsic alignment \citep{Pyne2021}, which would in turn likely produce gains in cosmological constraining power. Detailed quantification of these gains will require further analyses, which we leave for future work.

\section*{Acknowledgement}

We thank Alan Heavens for suggestions that led to some of the early tests in the paper and Elisabeth Krause for useful discussions. The computation presented here was performed on the High Performance Computing (HPC) resources supported by the University of Arizona TRIF, UITS, and Research, Innovation, and Impact (RII) and maintained by the UArizona Research Technologies department. SSB is supported by the Department of Energy Cosmic Frontier program, grant DE-SC0020215. ER's work is supported is supported by NSF grant 2009401.  ER also receives funding from DOE grant DE-SC0009913 and NSF grant 2206688.

\section*{Data availability statement}

The data underlying this article will be shared on request to the corresponding authors.
\bibliographystyle{mnras}
\bibliography{info_content} 


\bsp	
\label{lastpage}
\end{document}